\documentclass[fleqn,usenatbib]{mnras}
\usepackage{newtxtext,newtxmath}
\usepackage[T1]{fontenc}

\DeclareRobustCommand{\VAN}[3]{#2}
\let\VANthebibliography\thebibliography
\def\thebibliography{\DeclareRobustCommand{\VAN}[3]{##3}\VANthebibliography}

\usepackage{graphicx}	
\usepackage{amsmath}	
\usepackage[utf8]{inputenc}
\usepackage{indentfirst}
\usepackage{hyperref,graphicx} 
\usepackage{mathrsfs} 
\usepackage{amsmath,siunitx}
\usepackage{bm}
\graphicspath{{./fig/}}
\usepackage{natbib}
\usepackage[version=4]{mhchem}
\usepackage{lineno}

\newcommand{\co}{\ce{CO}}

\newcommand{\gd}{\Sigma_{\rm{gap}}/\Sigma_{0}}

\newcommand{\Me}{M_\oplus}
\newcommand{\Msun}{M_\odot}
\newcommand{\Mstar}{M_\star}
\newcommand{\Mp}{M_{\rm{p}}}
\newcommand{\Mj}{M_\mathrm{J}}

\newcommand{\rp}{r_{\mathrm{p}}} 
\newcommand{\Rstar}{R_\star}
\newcommand{\Rsun}{R_\odot}

\newcommand{\sigmag}{\Sigma_{\rm{gas}}}
\newcommand{\sigmagdl}{\Sigma_{\rm{gas}}/\Sigma_{\rm{gas,0}}}

\newcommand{\taur}{\tau_{\rm{rad}}}
\newcommand{\tauv}{\tau_{\rm{vert}}}
\newcommand{\Tit}{T_{\rm{iterate}}}
\newcommand{\Tmid}{T_{\rm{mid}}}
\newcommand{\Tstar}{T_\star}

\newcommand{\Tsub}{T_{\rm{sub}}}
\newcommand{\Tgas}{T_{\rm{gas}}}
\newcommand{\Tdust}{T_{\rm{dust}}}

\newcommand{\fargo}{\text{FARGO3D}}
\newcommand{\hd}{\text{hydrodynamical}}

\newcommand{\radmc}{\text{RADMC-3D}}
\newcommand{\rt}{\text{radiative transfer}}

\newcommand{\md}{\text{Model D}}  
\newcommand{\mg}{\text{Model G}}  
\defcitealias{chen_planet_2023}{C23}

\newcommand{\bt}[1]{\textnormal{#1}}
\newcommand{\btt}[1]{\textnormal{#1}}

\title[Substructure affects disk temperature]{Planet-induced Gas and Dust Substructure Feedbacks on Disk Thermal Structure}

\author[K. Chen et al.]{
Kan Chen,$^{1}$\thanks{E-mail: kan.chen.21@ucl.ac.uk}
Paola Pinilla$^{2}$
and Mihkel Kama,$^{1,3}$
\\
$^{1}$ Department of Physics and Astronomy, University College London, Gower Street, London, WC1E 6BT, UK\\
$^{2}$ Mullard Space Science Laboratory, University College London,
Holmbury St Mary, Dorking, Surrey RH5 6NT, UK\\
$^{3}$ Tartu Observatory, University of Tartu, Observatooriumi 1, Tõravere 61602, Tartu maakond, Estonia
}

\date{Accepted XXX. Received YYY; in original form ZZZ}

\pubyear{2025}

\begin{document}
\label{firstpage}
\pagerange{\pageref{firstpage}--\pageref{lastpage}}
\maketitle

\begin{abstract}
    
    \bt{Protoplanets can interact with their natal disks and generate gas and dust substructures such as gaps and rings. However, how these planet-induced substructures affect the disk temperature, and how that in turn influences the substructures, remains unclear. We aim to study disk substructures and the thermal structure self-consistently and explore their impact on volatile distribution. To this end, we perform iterative multi-fluid hydrodynamical and radiative transfer simulations of planet-disk interactions. We find that the temperature in a structured disk deviates significantly from that of a smooth disk due to giant planet formation. In particular, midplane temperatures in gaps can increase by tens of Kelvin, leading to volatile sublimation as well as radial shifts and multiplication of icelines. Comparing our multi-dust models with previous gas-only models, we find that the former produces slightly shallower gaps and temperatures about 10 K ($\sim25\%$) higher. Furthermore, the temperature at dust rings formed by pressure bumps can drop by several Kelvin, creating volatile freeze-out regions. Nevertheless, the overall midplane ice distribution is not strongly sensitive to whether dust is included. We also investigate the effect of varying disk viscosity. Increasing $\alpha$ viscosity from $10^{-4}$ to $10^{-2}$ leads to a roughly 10 K ($\sim25\%$) warmer midplane due to enhanced vertical dust mixing. However, higher viscosity suppresses gap opening and reduces the temperature enhancement within gaps. As a result, iceline locations do not follow a simple trend with viscosity. Finally, we propose an observational strategy using ALMA to test our predicted temperature changes within disk gaps.}


\end{abstract}

\begin{keywords}
protoplanetary discs -- planet-disc interactions -- hydrodynamics -- radiative transfer -- planets and satellites: composition
\end{keywords}



\section{Introduction}

High-resolution observations have revealed that protoplanetary disks (PPDs) are highly structured, displaying gaps, rings, spirals, and asymmetries \citep{andrews_disk_2018, long_gaps_2018, oberg_molecules_2021}. Some studies have attempted to measure disk temperatures in both radial and vertical directions using images and line profiles \citep{fedele_probing_2016, calahan_tw_2021}. Additionally, recent observations have reported non-smooth radial brightness temperature profiles, suggesting that gas and dust temperatures in disks may not be radially smooth \citep{lawMoleculesALMAPlanetforming2021, lawMoleculesALMAPlanetforming2021a, leemker_gas_2022, law_mapping_2024}. However, whether these disk substructures play a significant role in shaping the disk temperature remains an open question.

Monte Carlo Radiative Transfer (MCRT) simulations are used to model the disk temperature structure. Compared to other $\rt$ methods, MCRT has advantages in accurately handling dust opacities, absorption, scattering, and complex disk geometries. For example, by assuming a gap density profile induced by Jupiter in the solar nebula, \citet{turner_hot_2012} used MCRT to find an increase in gap temperature; while \citet{broome_iceline_2023} examined how the gap temperature deviates from that of a disk without a gap by considering stellar radiation and viscous heating, though they also assumed fixed gap density profiles. In both cases, the gas and dust density structures in the MCRT simulations were analytically prescribed and fixed in time.

In a more realistic simulation, substructures may alter the disk temperature, and changes in disk temperature may, in turn, feedback on disk substructures. For example, the gap-opening process is influenced by the disk scale height \citep{fung_how_2014, kanagawa_mass_2015, duffell_empirically_2020}, which is determined by the disk temperature. Another example is that temperature sets the locations of volatile icelines, which are crucial for planet and planetesimal formation \citep{oberg_effects_2011, schoonenberg_planetesimal_2017}. Icelines may lead to the formation of dust gaps or rings. Observations by \citet{zhang_evidence_2015} found that the dust continuum gaps in HL Tau align with several volatile iceline locations. Additionally, \citet{pinilla_dust_2017} used dust evolution simulations to show that icelines can induce rings or gaps in scatter light or dust continuum observations, depending on viscosity. However, later surveys \citep{huang_disk_2018, long_gaps_2018} found no simple one-to-one correlation between the radial locations of rings or gaps and expected iceline positions, assuming a smooth and monotonically decreasing radial temperature profile. These model and observation comparisons do not account for the possibility that a structured disk may have a structured temperature profile rather than a smooth one. Therefore, it is crucial to study disk temperature and disk substructures simultaneously and self-consistently.

Recently, \citet[hereafter \citetalias{chen_planet_2023}]{chen_planet_2023}
proposed a novel iterative hydrodynamical (HD) and MCRT method to study how a planet-induced gap can affect disk temperature. Simultaneously, the temperature change alters the disk scale height, further influencing the gap-opening process. They showed that in a gap induced by a Jovian planet at a few au (e.g., $4\,$au) to a few tens of au (e.g., $30\,$au), the midplane temperature can increase significantly, by up to several tens of Kelvin. As a result, volatiles such as \ce{CO} ice can sublimate, leading to multiple CO icelines and new iceline locations, which differ significantly from the number and position of icelines in a smooth disk. Furthermore, the complex iceline distribution suggests a complex C/O ratio across the disk, challenging the canonical C/O ratio derived from a smooth disk in \citet{oberg_effects_2011}. However, \citetalias{chen_planet_2023} did not account for dust dynamics during the iteration process.

In this paper, we aim to use HD-MCRT simulations with multiple dust species to investigate how dust and gas substructures generated by planet-disk interactions influence the disk temperature structure and volatile distribution. Studying gas and dust distributions simultaneously is crucial, as dust and gas interact with each other. Dust densities are affected by gas drag, which alters the dust distribution and dust-to-gas ratio \citep{weidenschilling_aerodynamics_1977}. Meanwhile, dust distribution and opacities strongly influence both dust and gas temperatures. These temperature changes, in turn, affect the gas density structure of the disk.

This paper is organized as follows. In Section\,\ref{sec:methods}, we describe our method of iterating HD and MCRT simulations with multiple dust species. In Section 3, we present our results, comparing our new method (which includes multiple dust species) with our previous approach (which did not) and exploring the effects of different viscosities using the new method. Section 4 discusses model simplifications and potential observational strategies to test our model. We summarize our findings in Section 5.

\section{Methods}\label{sec:methods}

We combine $\hd$ (HD) and Monte Carlo $\rt$ (MCRT) simulations and iterate them to study how substructures can affect disk thermal structures. We modify our previous iteration model in Figure 1 in \citetalias{chen_planet_2023} to include dust of different grain sizes in both hydrodynamics and radiative transfer. In our previous model, we did not include dust in HD simulations and simply assumed that $0.1\,\mu m$ dust was well mixed with gas in MCRT simulations.

Including multiple dust species is important: first, in the HD simulation, dust may decouple from the gas, depending on the Stokes number (St), which is defined as

\begin{equation}
    \mathrm{St} = \frac{\pi}{2} \frac{\rho_s a}{\Sigma_g},
\end{equation}
where $\rho_s$ is the internal density of the dust particle, $a$ is the dust grain size, and $\Sigma_g$ is the gas surface density. 

Large dust grains have larger St and can drift radially faster than small dust grains. Therefore, large grain sizes (e.g., 1 mm) can be distributed very differently from small grain sizes and gas. So we need to directly obtain dust density structures from HD simulations instead of assuming a well-mixed dust and gas density structure. 
Second, in MCRT, different grain sizes have different opacities and different levels of dust settling in disk vertical directions. Thus, the implementation of dust can alter the disk temperature, which can affect the planetary gap-opening process and further change the temperature at the gap as a consequence.

\begin{figure*}
\includegraphics[width=\linewidth]{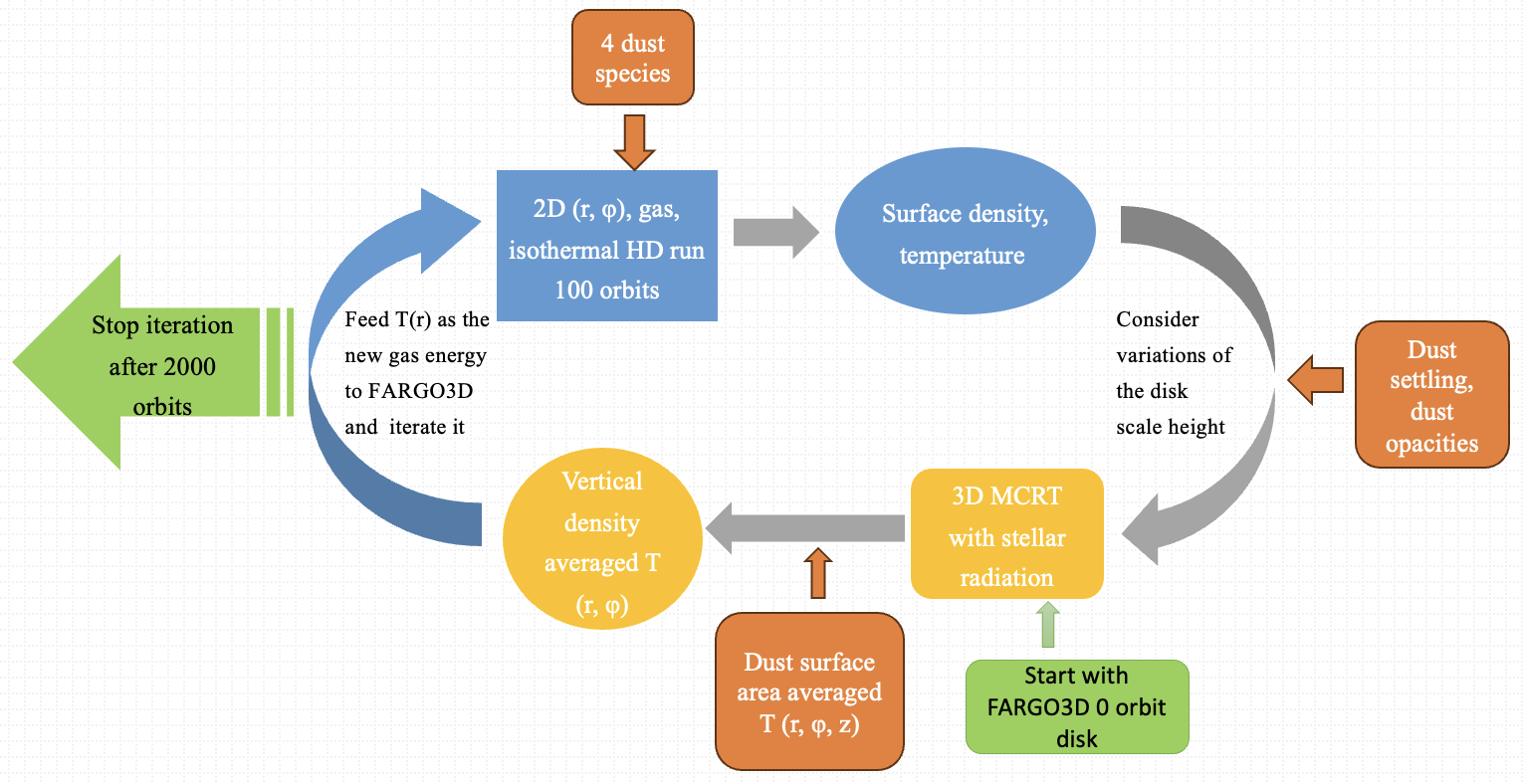}
\centering
\caption{Workflow of our iteration method with the implementation of multiple dust species. The workflow is modified from the workflow in \citetalias{chen_planet_2023} by adding multiple dust species in both HD and MCRT simulations.}
\label{fig:workflow}
\end{figure*}

We show our workflow for implementing multiple dust species into our iteration method in Fig. \ref{fig:workflow}. This workflow is modified from the workflow in \citetalias{chen_planet_2023} by adding multiple dust species in both HD and MCRT simulations. Basically, our iteration method can be understood as three main steps. First, we run HD simulations to get the density structure without temperature evolution. Second, we input the density from HD into MCRT simulations to get the temperature structure. Thus, there is no density evolution in MCRT.  \bt{Third, we process the MCRT \btt{multiple} dust temperatures to get gas temperature and feed it into the HD simulation \btt{(Section \ref{sec:rt2hd})}}. Then we repeat the first two steps above. We discuss more about these three steps in the following sections.
We perform the iteration process every 100 planetary orbits and iterate to 2000 orbits in total. We also carry out parameter studies of different planet masses $\Mp$, planet locaitons $\rp$, and $\alpha$ viscosities (shown in Table \ref{tab:fg}).

\subsection{Hydrodynamical setups}
For the setup of HD simulations, Table \ref{tab:fg} summarizes our parameter space. We use the same disk and planet setup as \citetalias{chen_planet_2023}. The entire disk extends from 1 to 100 au. We fix the planet on a circuler orbit.
We run multi-species $\fargo$ simulations with gas plus four different grain sizes \citep{weber_predicting_2019} in 2D, radial and azimuthal directions (r, $\phi$). Readers are referred to the governing equations in Section 2.1 of \citet{weber_predicting_2019}. We use four grain sizes $a_i$ of 0.1 $\mu m$, 2.2$ \mu m$, 46 $\mu m$, and 1 mm and the number density of grain sizes follow a power law $n(a) \propto a^{-\gamma_{\text{dust}}}, \gamma_{\text{dust}}=3.5$.
The internal density of grains is $3.7 g/cm^{3}$. For simplification, we refer to our previous model with only gas in hydro simulations and with 0.1 $\mu m$ dust in MCRT simulations as "Model G" and our new model with multiple dust species as "Model D". We include dust feedback on the gas, as well as dust diffusion, in our simulations.

Additionally, we study the effect of different viscosities in $\md$. The level of disk turbulence can affect the HD simulations in several ways, including gap opening and dust diffusion. Also in \btt{the 3D density setup of} MCRT simulations, different $\alpha$ \bt{impacts turbulent mixing, subsequently affecting} dust settling. Inspired by recent turbulence measurements in several disks \citep{flaherty_weak_2015, pinte_dust_2016, flaherty_turbulence_2018, teague_temperature_2018, dullemond_disk_2018, flaherty_measuring_2020}, we adopt $\alpha = 10^{-2}$, $10^{-3}$, and $10^{-4}$.

\begin{table}
\caption {$\fargo$ main parameters. Parameters in each column below the nineth row in this table are corresponding to the cases of planet location $\rp=$ 4, 10, or 30 au, respectively. Different $\alpha$ studies are only for cases of 100$\mathrm{M}_{\oplus}$.}
\label{tab:fg} 
\centering
\begin{tabular}{llll}
\hline
\hline
parameters      & \multicolumn{3}{c}{values}     \\
\hline
$\Mp$  & \multicolumn{3}{c}{{3$\mathrm{M}_{\mathrm{J}}$, 100$\mathrm{M}_{\oplus}$, 10$\mathrm{M}_{\oplus}$}}       \\
$a_i$[$\mu m$]    &  \multicolumn{3}{c}{0.1, 2.2, 46, 1000} \\
SigmaSlope      & \multicolumn{3}{c}{-1}          \\
FlaringIndex    & \multicolumn{3}{c}{0.25}       \\
$M_{\star}$ [$M_{\odot}$] & \multicolumn{3}{c}{1.0} \\
$\rho_{solid}$ [g/cm$^3$]              & \multicolumn{3}{c}{3.7} \\
$\gamma_{\text{dust}}$ & \multicolumn{3}{c}{-3.5} \\
$\epsilon$                 & \multicolumn{3}{c}{0.01} \\
\hline
$r_{0}$ = $r_p$ {[}au{]} & 4        & 10       & 30       \\
$r_{min}$ {[}$r_{0}${]}      & 0.25     & 0.1      & 0.033    \\
$r_{max}$ {[}$r_{0}${]}      & 25       & 10       & 3.3      \\
Aspect Ratio     & 0.04     & 0.05     & 0.066    \\
$\Sigma_{0}[\Mstar/r_{0}^2]$      & $1.8^{-4}$ & $4.5^{-4}$ & $1.34^{-3}$ \\
$N_{r, HD}$               & 580      & 460      & 350      \\
$N_{\phi, HD}$            & 790      & 630      & 480      \\
\hline
$\alpha$ (for 100$\mathrm{M}_{\oplus}$)           & \multicolumn{3}{c}{$10^{-2}, 10^{-3}, 10^{-4}$ }      \\
\hline
\end{tabular}
\end{table}

For boundary conditions, we use a damping prescription to minimize reflections near the boundaries. Similar to \citet{pyerin_constraining_2021}, for the radial boundaries, we adopt power-law extrapolation for densities and Keplerian extrapolation for azimuthal velocities for both dust and gas. Regarding the radial velocities for gas and dust, we adopt symmetric inner and outer boundaries to conserve the disk mass. Periodic boundaries are imposed in the azimuthal direction.

\subsection{Monte Carlo Radiative Transfer setup}
We use the 3D Monte Carlo radiative transfer code, RADMC-3D \citep{dullemond_radmc-3d_2012}, to calculate the disk temperature structure. For the setup of the MCRT, we summarize our parameter choices in Table \ref{tab:rad}, which are mainly similar to the parameter choices in \citetalias{chen_planet_2023}. 
\btt{We use the same radial simulation domain as in the HD simulations, from 1 to 100 au.}
In this project, we azimuthally average the model to minimize the effect of poor photon statistics, except for cases developing eccentric gaps, such as 3$\Mj$ at 4 au and 3$\Mj$ at 10 au with $\alpha = 10^{-3}$ and $\alpha = 10^{-4}$, respectively. 
The star and disk setup are the same as the counterparts in \citetalias{chen_planet_2023} \bt{except for} the treatment of dust which we discuss below.

As we have four dust species, ranging from 0.1 $\mu m$ to 1 mm, we calculate the corresponding dust opacity for different grain sizes using the optool package \citep{dominik_optool_2021}. We input the dust surface densities and  the evolving gas scale height from HD simulations into MCRT simulations. Following equation (19) in \citet{fromang_global_2009}, the vertical distribution of the dust is calculated from gas scale height, $\alpha$ and St.
This is a steady state dust vertical distribution when turbulent diffusion balances dust settling. 
In other words, a larger grain size or a smaller $\alpha$ viscosity leads to a smaller dust scale height. As a result, we can extend the 2D dust surface density into a 3D dust column density. We consider isotropic scattering for dust.

\subsection{Prossessing between radiative transfer and hydro} 
\label{sec:rt2hd}

From the MCRT in $\md$, we obtain the dust temperature for each grain size in each grid cell, $T_{dust, i}(r, \phi, z)$, where $i$ represents the $i$-th dust species. However, in the HD simulations, we need the gas temperature in the HD grid cell ($r, \phi$). Therefore, when feeding the MCRT temperature to the HD simulations, we carry out the following processes.

First, we calculate the dust surface-area-averaged temperature in each grid cell, $T_{dust}(r, \phi, z)$.

\begin{equation}
T_{dust}(r,\ \phi ,\ z) =\frac{\sum\limits _{i} A_{i} \ n_{i}( r,\ \phi ,\ z) \ T_{dust,i}( r,\ \phi ,\ z )}{\sum\limits _{i} A_{i\ } n_{i}( r,\ \phi ,\ z )} \ \ 
\label{eq:tdust_av}
\end{equation}
where $n_{i}$ is the dust number density of species $i$, $A_{i}=4 \pi a_{i}^{2}$, $i$=1,2,3,4. Note that for $\mg$, since we only have one dust species, we do not need to do the above averaging.

Second, the surface-area-averaged dust temperature is a good approximation for the gas temperature when gas and dust particles are well mixed \citep{facchini_different_2017}. Therefore, we assume $T_{dust}(r, \phi, z) = T_{gas}(r, \phi, z)$.

Third, we calculate a vertically-averaged density-weighted gas temperature $\bar{T}_{gas}(r, \phi)$ from $T_{gas}(r, \phi, z)$ by using

\begin{equation}
\bar{T}_{gas}(r, \phi)=\frac{\int T_{gas} (r,  \phi, z) \rho_{gas}(r, \phi, z) dz}{\int \rho_{gas}(r, \phi, z) dz}.
\end{equation}

\bt{
Fourth, we input $\bar{T}_{gas}(r, \phi)$ as the gas temperature for the next round of 2D HD run. Note that as our simulations combine HD and MCRT simulations for each iteration step (100 orbits) and iterate them until 2000 orbits for a steady state, we also refer to $\bar{T}_{gas}(r, \phi)$ at the final 2000 orbit as the iterative temperature $\Tit$ for simplicity in the following sections.
}

\begin{table}
\centering
\caption{$\radmc$ parameters.}
\label{tab:rad}
\begin{tabular}{ll}
\hline
\hline
parameters   & values \\
\hline
$\Mstar$ [$\Msun$] & 1      \\
$\Rstar$ [$\Rsun$] & 1.7    \\
$\Tstar$ [K]    & 4730   \\
$N_{photon}$      & $10^8$   \\
$N_{r, MCRT}$            & 256    \\
$N_{\phi, MCRT}$         & 1     \\
$N_{\theta, MCRT}$     & 53    \\
\hline
\end{tabular}
\end{table}

\section{Results}

We compare the results of our iteration methods with and without multiple dust species. To this end, we present the results of general disk modeling, including density structure, temperature structure, and iceline locations.

\subsection{Effects of dust}

In this section, we present and analyze the results obtained from $\mg$ in \citetalias{chen_planet_2023} and $\md$ (this work) with the same viscosity $\alpha=10^{-3}$.

\subsubsection{Density maps}

\begin{figure*}
\includegraphics[width=\linewidth]{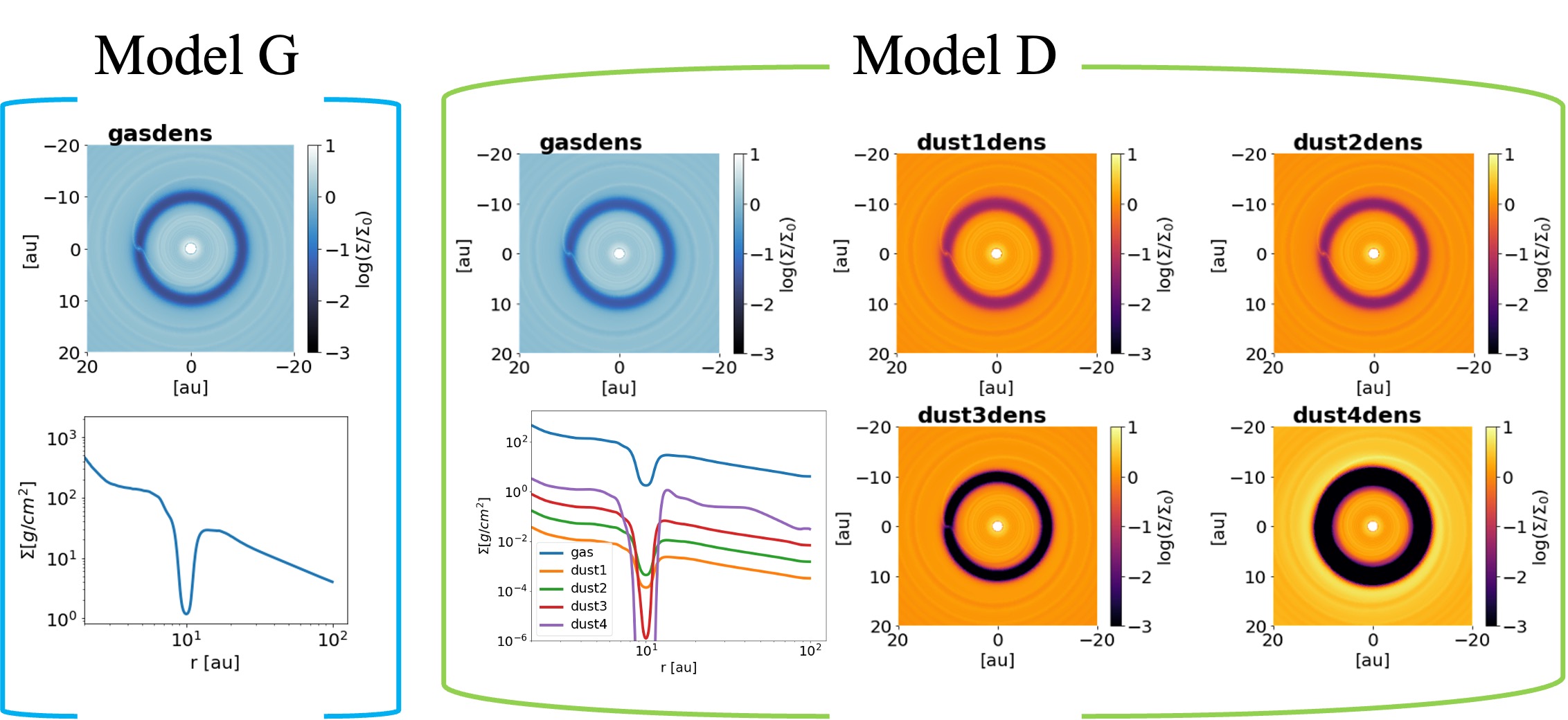} 
\centering
\caption{Comparisons of surface density between the $\mg$ (left) and $\md$ (right) with iterations for 100$\Me$ at 10au at 2000 orbits. For four dust species in $\md$, with panels named from dust1dens to dust4dens, they are $0.1 \mu m, 2.2 \mu m, 46 \mu m$ and 1mm, respectively. 2D surface density maps are shown in units of normalized densities, while the 1D radial surface density profiles are shown in absolute values.
}
\label{fig:sigma_100me10au} 
\end{figure*}

We show the surface densities of 100$\Me$ at 10au at 2000 planetary orbits from the $\mg$ (left) and $\md$ (right) in Fig.~\ref{fig:sigma_100me10au}. For the gas surface density, both models show similar simulation features in the 2D surface density map, such as spirals and concentric gaps. For dust densities in $\md$, the 2D density maps show that dust gaps across four different grain sizes are concentric with 100$\Me$ at 10au. Similar to \citet{rosotti_minimum_2016}, as the grain size increases, the dust gap becomes deeper and wider (as shown in the 1D radial profile in $\md$). Specifically, the gap in 1mm dust is about 5 au (0.5 $\rp$) wide, which is roughly 2 times wider than the 0.1 $\mu m$ one. Additionally, the former could be very depleted ($<10^{-6} g/cm^{-2}$) in the gap center, while the latter is only about 10 times lower than its initial values. This is expected, as the small grains couple well with the gas and follow the gas distribution, while large grains either experience radial drift inward inside the gap or get trapped in pressure maxima outside the gap. Thus, large grains are prevented from filling the gap.

\subsubsection{Radial profiles of density, aspect ratio and temperature}
\label{sec:3plot}

\begin{figure*}
\includegraphics[width=\linewidth]{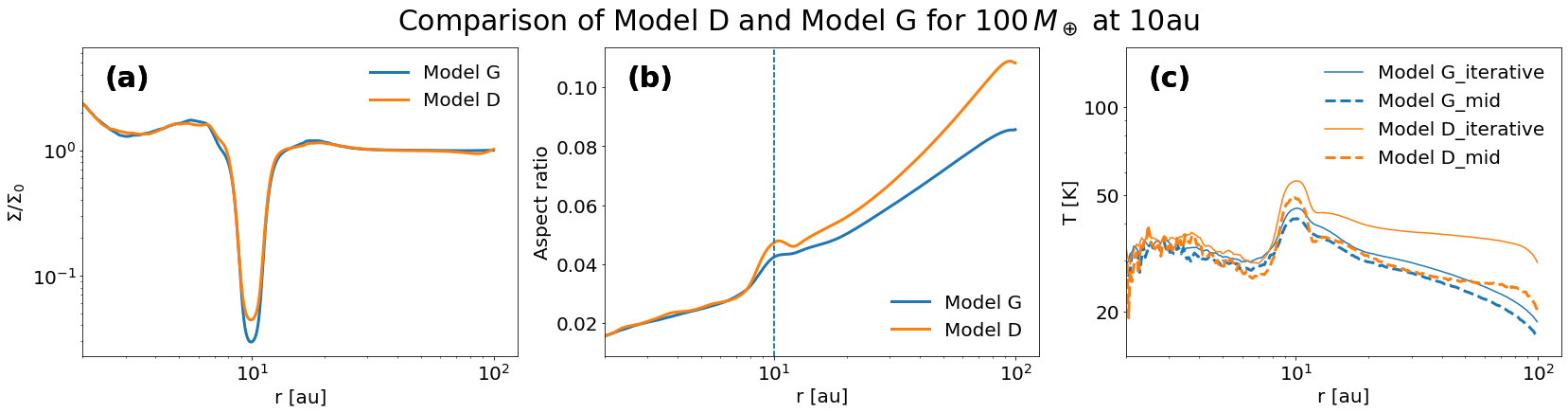} 
\centering
\caption{Radial profiles of gas surface density (a), gas aspect ratio (b), and temperature (c) of iterative process $\Tit$ (solid lines) and midplane $\Tmid$ (dashed lines) of 100$\Me$ at 10au at 2000 orbits obtained from $\mg$ and $\md$, respectively.
} 
\label{fig:3plot_100me10au} 
\end{figure*}

We show the comparison of the gas surface density, disk aspect ratio, and disk iterative and midplane temperature as a function of disk radius of 100$\Me$ at 10au at 2000 orbits obtained from $\mg$ and $\md$ in Fig.~\ref{fig:3plot_100me10au}. We remind the reader that, as explained in Section \ref{sec:rt2hd}, the iterative temperature $\Tit$ is the temperature used for each HD step. It is a vertically-averaged, density-weighted temperature. Another temperature we are concerned about is the midplane temperature $\Tmid$ because ices of volatiles mainly locate at the midplane.

In the gap region, we find that the gas gap is shallower when calculated by $\md$ than by the $\mg$. The former is $\sigmagdl = 2.5\times10^{-2}$ at the gap center, while the latter is $4\times10^{-2}$. This is explained by the iterative temperature $\Tit$ at the gap being about 10\,K \bt{($\sim$ 20\%)} higher in $\md$ than in $\mg$, which makes the gas aspect ratio slightly increase from 0.045 ($\mg$) to 0.05 ($\md$). A higher aspect ratio makes gap opening more difficult \citep{crida_width_2006}.  We also find that $\Tmid$ is about 10\,K \bt{($\sim$ 25\%)} higher in $\md$ than in $\mg$ in the gap region.

In the regions outside the gap (Fig. \ref{fig:3plot_100me10au}(c)), $\md$ presents a similar $\Tit$ as $\mg$ for $r < \rp$.
We also note that $\Tit$ is approximately 10\,K higher in $\md$ than in $\mg$ in regions outside the gap, $r > \rp$, whereas the difference in $\Tmid$ is smaller.

Inside $\md$, $\Tit$ at the gap center is about 55\,K, which is about 10\,K higher than the outer gap edge (14 au) and about 25\,K higher than the inner gap edge (7 au). The inner gap edge is cooler because the inner dust rim at 1 au is puffed up, creating a shadowed region that extends up to about 10 au. \bt{This shadowing effect is also discussed with more details in Section 4.3 in \citetalias{chen_planet_2023}.} 

We check the gap opening in other $\Mp$ and $\rp$ cases and find that they show the same trend as 100$\Me$ at 10au, where the gap depth in $\md$ is shallower than that in $\mg$. There are some exceptions for cases with the presence of eccentric gaps caused by $3\Mj$. In that case, the gas gap depth of $\mg$ is shallower than that in $\md$. This is because the more eccentric gap in $\mg$ creates a denser streamer, which enhances the azimuthal average density at the gap.

\begin{figure*}
\includegraphics[width=\linewidth]{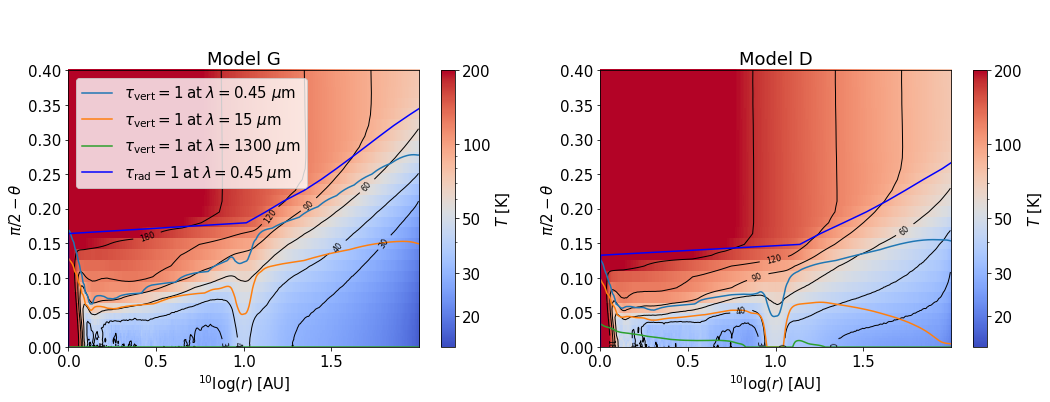} 
\centering
\caption{$\tau=1$ surfaces at differen wavelengths of $\mg$ (left) and $\md$ (right) of 100$\Me$ at 10au at 2000 orbit, respectively. Background colormap is the dust temperature in $\mg$ and dust surface area averaged temperature in $\md$. Vertical and radial $\tau=1$ surfaces at different wavelengths are shown ln lines with different colors. \bt{We mark isothermal contours of 30, 40, 60, 90, 120, and 180 K (black solid lines) in the temperature maps. We also show a residual plot and a ratio plot between these two temperature maps for better visualing the temperature differences
in Appendix \ref{sec:app_Tdiff_modelg_modeld}}.
}
\label{fig:tau1_wi_wt_dust} 
\end{figure*}

To better understand the temperature differences between $\mg$ and $\md$, we show in Fig.\,\ref{fig:tau1_wi_wt_dust} a 2D cut of the temperature structure with $\tau=1$ surfaces at different wavelengths.
Dust absorbs stellar photons at short wavelengths (optical, NIR; $\lambda=0.45\,\mu$m is shown for reference) and re-emits at longer wavelengths (mid-IR to mm; $\lambda=15\,\mu$m and $1300\,\mu$m).
In $\mg$, the radial $\taur = 1$ and vertical $\tauv = 1$ optically thick surfaces at $\lambda = 0.45\,\mu$m are higher, meaning that fewer photons at the stellar intensity peak wavelength can penetrate into the disk. In contrast, at $\lambda = 1.3$\,mm, $\mg$ is optically thin, whereas $\md$ is \bt{marginally} optically thick. This results from the larger long-wavelength opacity of mm-grains, which are absent in $\mg$. The slightly lower opacity makes it easy for cooling radiation to escape in $\mg$.
\bt{As a combined effect, $\mg$ generally has a lower temperature (about 30\% lower) than $\md$ at the same grid cell in most regions except the disk midplane.
Nevertheless, it is important to note that for the midplane temperature $\Tmid$, the difference between $\mg$ and $\md$ is relatively small (within 15\%).}
On the other hand, the difference between $\mg$ and $\md$ is larger in the vertically integrated, density-weighted temperature $\Tit$. 
\bt{For a direct temperature comparison, we also refer to check Appendix \ref{sec:app_Tdiff_modelg_modeld}.}

\bt{In addition, we test a case of 100$\Me$ at 10 au with double the number of grain sizes between 0.1~$\mu$m and 1~mm (with grain size distribution still following the -3.5 power law) for our iteration modeling. We find that the temperature difference at 2000 orbits between the 8-dust-species model and the 4-dust-species model is low (less than 15\%) in most
regions of the disk. Therefore, we consider the 4-dust-species model sufficient for our purposes.}

\subsubsection{Temperature at dust rings}

We also study how dust rings, formed by dust trapping at pressure maxima, can affect disk temperature.

For the case of 3$\Mj$ at 30au in $\md$, a strong mm dust-trap ring is present at the outer edge of the gap \citep{pinilla_trapping_2012, pinilla_ring_2012} at $r \sim 50-60$ au. In contrast, no dust ring is present in $\mg$. This is because $\mg$ only considers 0.1 $\mu m$ dust, which is assumed to follow the gas density, and there is no obvious gas ring at the outer gap edge.

\begin{figure}
\includegraphics[width=\columnwidth]{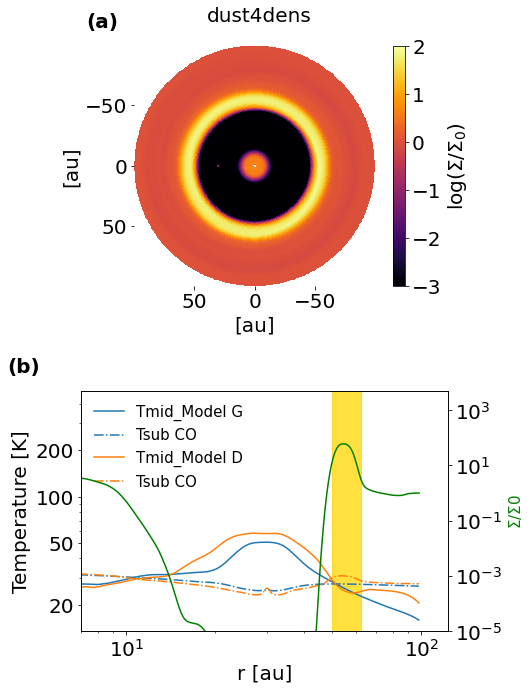} 
\caption{Dust trap in mm size grains of 3$\Mj$ at 30au at 2000 orbits. Panel (a) shows surface density map of mm dust and panel (b) shows \btt{midplane temperature $\Tmid$ (solid lines) and sublimation temperature $\Tsub$ (dashed lines)} from $\mg$ and $\md$. In panel (b), the green line and the right-hand-side axis represent normalized mm dust density radial profile in $\md$ and the gold-shaded radius regions correspond to the mm dust ring regions in the density map.
}
\label{fig:dust_trap} 
\end{figure}

Figure \ref{fig:dust_trap} shows a 2D mm dust density map in $\md$ (panel (a)) and the radial midplane and sublimation temperature \btt{$\Tsub$ (the temperature at which ice turns directly into gas, \citealt{hollenbach_water_2009})} profiles (panel (b)) obtained from $\mg$ and $\md$. At the dust trap location in Fig. \ref{fig:dust_trap}(a), the mm dust surface density increases by more than 100 times from the initial condition, leading to a higher optical depth. As a result, the mm dust ring induces an approximately 5\,K drop \bt{(from about 30\,K decreases to 25\,K)} in $\Tmid$ compared to cases without a dust trap at the same radius in $\md$, though the former is still nearly the same as $\Tmid$ in $\mg$. However, this temperature drop at the dust ring is small and likely difficult to detect in ALMA observations.
We note that \citet{zhang_self-consistent_2021} also found a disk temperature drop at the dust ring, although they directly assumed a Gaussian density profile for the width and peak of the dust ring.

To confirm whether the temperature drop in the dust rings is due to the higher optical depth of the dust ring or the shadowing effect caused by the puffed-up disk scale height at the gap (with temperature enhancement), we performed a test using a puffed-up disk model without dust surface density enhancement at the previous dust ring location. We found that the temperature change at the dust ring location was negligible. In other words, the temperature drop at the dust ring is indeed primarily due to the optical depth effect of the dust ring.

Meanwhile, \bt{at the dust ring in Fig.\ref{fig:dust_trap}(b),} $\Tsub$ is about 5\,K higher in $\md$ than in $\mg$, while the $\co$ iceline shifts only slightly, moving a few au closer to the star. As $\Tmid$ is lower than the $\Tsub$ of $\co$ at the dust ring, the dust ring can act as a volatile freeze-out region, similar to the findings in \citet{alarcon_chemical_2020}. In addition, the hot gap can act as a highly active volatile sublimation region. Consequently, a significant amount of gas-phase volatiles diffuse to the outer gap edge and freeze out at the dust ring. This combination of a hot gap and a cold ring could provide a favorable environment for efficient pebble and planetesimal growth.

\subsubsection{Effects on icelines}

\begin{figure}
    \includegraphics[width=0.9\linewidth]{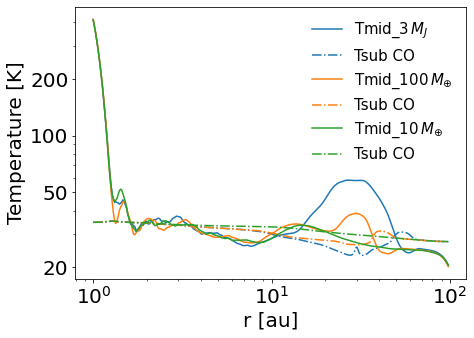} 
    \centering
    \caption{Miplane (solid) and sublimation (dashed) temperature profiles of differemt $\Mp$ at $\rp=$ 30 au of $\md$}
    \label{fig:tmid_rp30} 
\end{figure}

\begin{figure*}
\includegraphics[width=\linewidth]{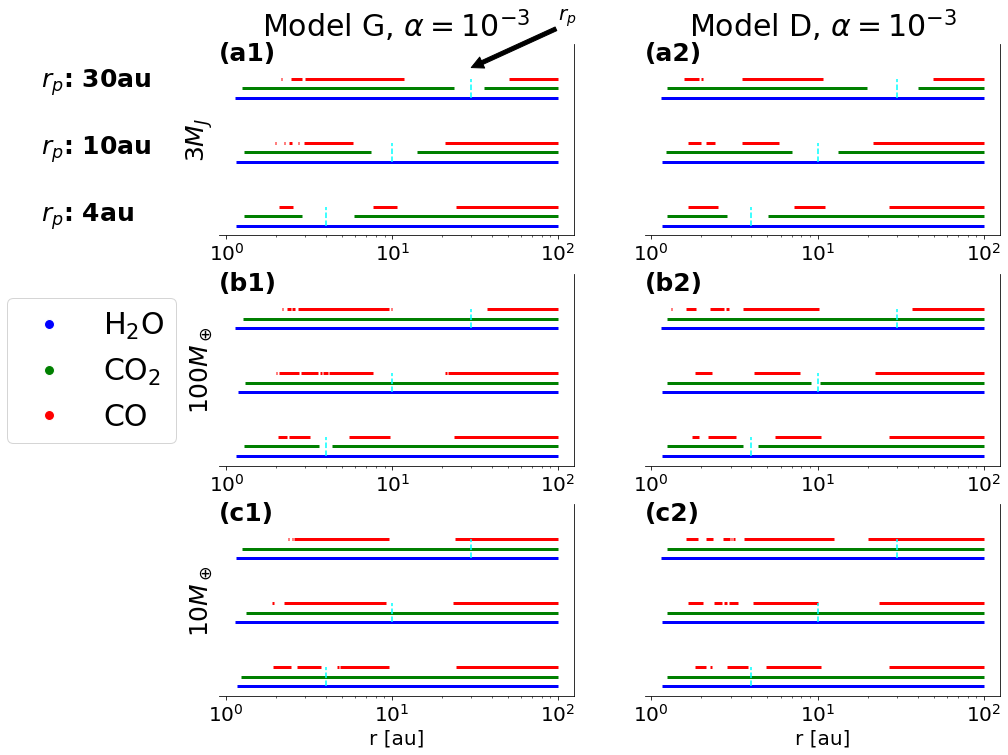} 
\centering
\caption{Comparions of radial ice distribution of \ce{H2O}, \ce{CO2} and \ce{CO} obtained from $\mg$ (left) and $\md$ (right). We show different $\Mp$, $3\Mj$, $100\Me$ and $10\Me$ from top to bottom. In each panel, from top to bottom, $\rp$ is 4, 10, and 30 au, respectively. Each bar represents ice existance regions in the midplane. Vertical cyan dashed lines represent the planet location. 
}
\label{fig:ice_dust} 
\end{figure*}

In Fig.\,\ref{fig:tmid_rp30}, we show the midplane and sublimation temperature for different planet masses ($\Mp$) at $\rp=30\,$au for the \md\ case. Since $\Mp=10\Me$ is unable to open a deep gap, the temperature profile is similar to the smooth disk. As $\Mp$ increases, the gap region becomes hotter, which is consistent with the results in \citetalias{chen_planet_2023}. 
$\Tmid$ is about 25\,K at $\rp$.
The midplane temperature of 3$\Mj$ at 30 au is about 20\,K higher than that of 100$\Me$ at 30 au, and the latter is about 10\,K higher than that of 10$\Me$ at 30 au.
The sublimation temperature of volatiles (e.g. CO) gets lower in gaps as $\Mp$ increases. Overall, the temperature contrast between gap and outside gap shows that a Saturn mass or even more massive planet can significant change the disk temperature structure.

As the midplane temperature of the disk and the pressure-dependent sublimation temperature of volatiles can be obtained from our models, we can proceed to investigate the behaviour of volatile icelines.
Figure\, \ref{fig:ice_dust} shows a comparison of the radial H$_{2}$O, CO$_{2}$, and CO ice distribution for the case of $\mg$ and $\md$.

Overall, the iceline locations from these two models are similar. This is because we mainly focus on the ice distribution in the midplane, and the $\Tmid$ values from $\mg$ and $\md$ are not significantly different, as shown in Fig. \ref{fig:3plot_100me10au}\,(panel c). Specifically, beyond 10 au, the number of icelines for a given volatile is essentially the same in both $\mg$ and $\md$, though their exact locations may differ.

The main differences in the gap regions appear in the case of a 3 $\Mj$ at 30 au, where $\md$ exhibits a wider \ce{CO2} sublimation region (green bars are \ce{CO2} ice region) around the planet (marked by vertical cyan dashed lines) compared to $\mg$. In $\md$ (the top third of panel (a2)), the \ce{CO2} sublimation region extends from 20 to 40 au. In contrast, in $\mg$ (the top third of panel (a1)), the \ce{CO2} sublimation region ranges from 25 to 35 au. This difference arises because $\Tmid$ is higher in the gap regions of $\md$ than in $\mg$.

Outside the gap regions, the main differences occur in the inner disk, within the first few au. Although the $\Tmid$ values from $\mg$ and $\md$ in this region differ by only a few Kelvin, the $\Tmid$ here is very close to the sublimation temperature of \ce{CO}. As a result, the \ce{CO} iceline is highly sensitive to even small differences in $\Tmid$ between $\mg$ and $\md$. We have more discussions about these compact ice regions in first few au in Section \ref{sec:flickering_ice}. \bt{But again, as pointed out in Section \ref{sec:3plot}, the first few au are shadowed by the inner rim set at 1 au. The shadowing effect causes the temperature drop, thus allowing \ce{CO2} and \ce{CO} ice in this area. However, the first few au may reach higher temperatures, and the \ce{CO} ice regions may disappear if we set the inner rim closer to the star. Instead, some \ce{CO2} ice regions might remain.}

\subsection{Effects of different viscosity}

\bt{
Different levels of turbulent viscosity can alter dust settling by affecting turbulent mixing, which in turn influences the disk temperature. Specifically, the vertical spreading of dust grains occurs through dust diffusion, given by $ D = \nu / Sc $, where $ \nu $ is the turbulent viscosity and $ Sc $ is the Schmidt number \citep{dullemond_effect_2004}. Typically, $ Sc $ is assumed to be 1. Therefore, higher viscosity leads to stronger dust diffusion, reduced dust settling, and an increased dust scale height. This effect can be captured by using equation (19) from \citet{fromang_global_2009} to set up the dust distribution in the MCRT simulations.
At the same time, viscosity also influences both the gap-opening process and dust diffusion in the HD simulations.   
}

\subsubsection{Disks without planets} \label{sec:visc_noplanet}

\begin{figure}
\includegraphics[width=0.9\columnwidth]{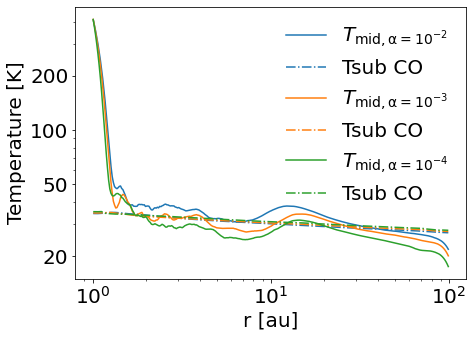} 
\centering
\caption{
Radial midplane temperature (solid lines) and sublimation temperature (dashed lines) of $\md$ in non-planet disks with $\alpha= 10^{-2}, 10^{-3}$ and $10^{-4}$, respectively. 
}
\label{fig:tmid_visc_noplanet} 
\end{figure}

\begin{figure*}
\includegraphics[width=\linewidth]{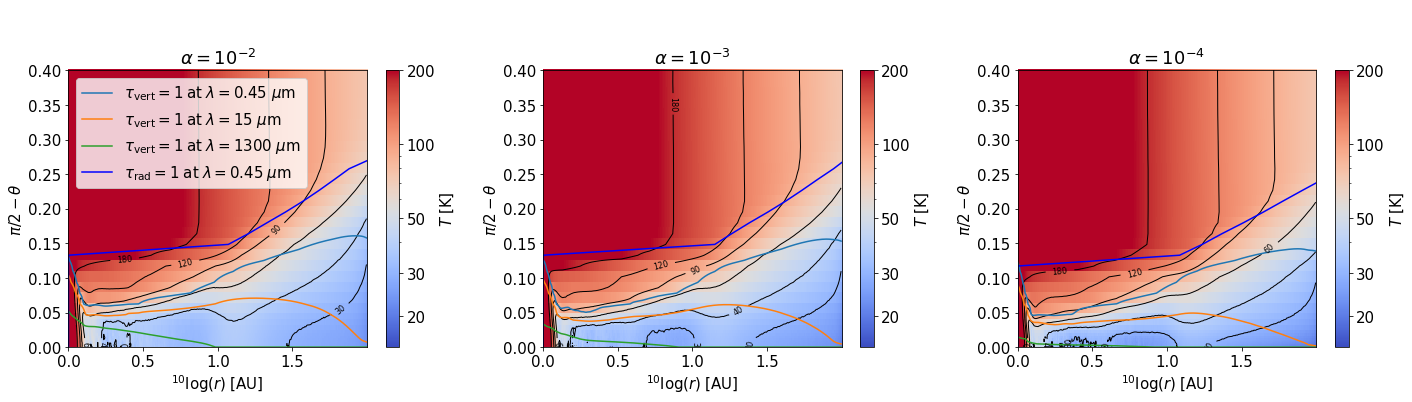} 
\centering
\caption{
Similar to Fig. \ref{fig:tau1_wi_wt_dust}, but for a planet-free disk setup with different viscosities in $\md$. 
\bt{From  left to right, $\alpha= 10^{-2}$, $10^{-3}$ and $10^{-4}$. We also provide a residual plot and a ratio plot between the temperature maps of $\alpha=10^{-2}$ and $10^{-4}$ for better visualizing the temperature differences in Appendix \ref{sec:app_Tdiff_visc}}.
}
\label{fig:tau1_visc_noplanet} 
\end{figure*}

When there is no planet in a disk, different turbulence levels can affect \bt{turbulent mixing then affect} dust settling, which in turn changes the height of the $\tau=1$ surface and impacts the disk's temperature distribution.
Figure~\ref{fig:tmid_visc_noplanet} compares the midplane temperature, $\Tmid$, as a function of radius for non-planetary disks with viscosities of $\alpha = 10^{-2}, 10^{-3},$ and $10^{-4}$. As $\alpha$ decreases, $\Tmid$ becomes lower across the entire disk. The $\Tmid$ for $\alpha=10^{-2}$ is about 10\,K \bt{($\sim 30\%$)} higher than that for $\alpha=10^{-4}$ at $r \leq 10$ au and a few K \bt{($\sim 20\%$)} higher at larger radii.

\bt{To explain the temperature difference in non-planetary disks with different viscosities, in Fig.~\ref{fig:tau1_visc_noplanet}, we show vertical slices of the temperature maps, as well as their $\tau=1$ surfaces at different wavelengths. Overall, the temperature is cooler in the disk surface but warmer in the midplane as $\alpha$ decreases.
More specifically, consider the cases of $\alpha=10^{-2}$ and $\alpha=10^{-4}$ for comparison (also see Fig.~\ref{fig:Tdiff_visc}). In the disk surface (roughly along the $\taur=1$ surface at 0.45~$\mu$m), the higher $\alpha$ results in stronger turbulent mixing, leading to a higher $\taur=1$ at 0.45~$\mu$m. Because dust located below the $\taur=1$ surface at 0.45~$\mu$m is not directly exposed to starlight, its temperature becomes lower. As a result, stellar photons are less able to penetrate deeply in the case of $\alpha=10^{-2}$ compared to $\alpha=10^{-4}$, making the temperature of $\alpha=10^{-2}$ about 15\% lower.
In contrast, near the disk midplane, a higher $\alpha$ leads to a higher $\tauv=1$ at 1.3mm, which reduces the cooling efficiency of the disk. Therefore, the midplane temperature of $\alpha=10^{-2}$ is about 15\% to 30\% higher than that of $\alpha=10^{-4}$.
}

Note that our radiative transfer simulations do not include viscous heating, we further discuss this effect in Section \ref{sec:limits}.
Additionally, as shown in Fig.~\ref{fig:tmid_visc_noplanet}, the sublimation temperature of a specific volatile is not strongly influenced by variations in viscosity.

\begin{figure}
\includegraphics[width=\columnwidth]{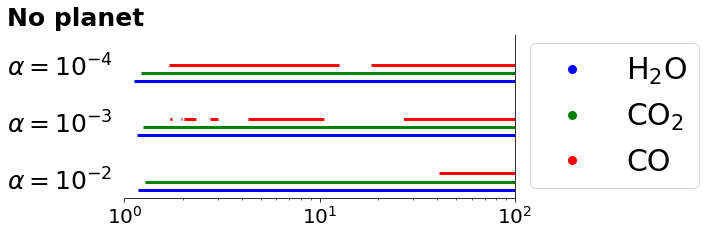} 
\centering
\caption{ 
Ice plot with no planets in disks with different viscosities. 
}
\label{fig:ice_visc_noplanet} 
\end{figure}

As a result, Fig.~\ref{fig:ice_visc_noplanet} presents the ice distribution in a non-planetary disk with different $\alpha$ values. The \ce{H2O} and \ce{CO2} icelines are located at approximately 1.15 and 1.3 au, respectively, when $\alpha = 10^{-2}$. These icelines shift slightly inward by about 0.05 au for \ce{H2O} and 0.15 au for \ce{CO2} when $\alpha$ decreases from $10^{-2}$ to $10^{-4}$.

The most significant iceline shift occurs for \ce{CO}. In the $\alpha = 10^{-2}$ model, the \ce{CO} iceline is located at $r \simeq 40$ au. However, for $\alpha = 10^{-3}$, multiple \ce{CO} icelines appear at $r \leq 5$ au. This phenomenon arises because, in the shadowed region of the disk (within the first few au), the midplane temperature $\Tmid$ for $\alpha = 10^{-3}$ coincidentally approaches the \ce{CO} sublimation temperature $\Tsub$ (as shown in Fig.~\ref{fig:tmid_visc_noplanet}). This intriguing scenario suggests that, under specific disk conditions related to density and turbulence levels, a non-planetary disk could host multiple icelines for one specific volatile within its self-shadowed region.
For $\alpha = 10^{-4}$, the inner \ce{CO} iceline is located at $r \simeq 2$ au. Additionally, we identify a region devoid of \ce{CO} ice between 10–30 au for $\alpha = 10^{-3}$ and between 10–20 au for $\alpha = 10^{-4}$. This occurs because these regions lie just outside the disk's self-shadowed area, allowing \ce{CO} ice to sublimate before freezing out again at larger disk radii.

\subsubsection{Disks with a 100 $\Me$ planet}

\begin{figure*}
\includegraphics[width=\linewidth]{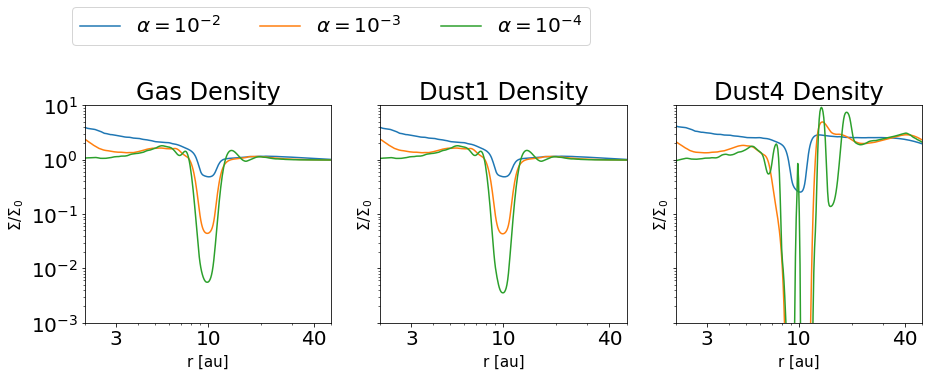} 
\centering
\caption{
The surface density of gas, 0.1 $\mu m$ and 1mm dust (left to right) as a function of disk radius of 100$\Me$ at 10au at different viscosities. The surface density is normalized by the initial value.
}
\label{fig:sigma_visc} 
\end{figure*}

In our simulations, $\Mp \geq 100\Me$ can open deep gas gaps ($\sigmag/ \Sigma_{gas,0} < 0.1$) in all cases except when $\rp = 10, 30$ au at $\alpha = 10^{-2}$ and $\rp = 30$ au at $\alpha = 10^{-3}$.
Figure \ref{fig:sigma_visc} shows the normalized radial surface density profiles of gas and dust across the disks at different $\alpha$ viscosities for a 100 $\Me$ planet at 10 au. Two representative dust sizes, 0.1 $\mu$m and 1 mm, are shown. Overall, as viscosity decreases, the gas and dust in the disks become more structured. The locations of gaps and rings in the dust generally coincide with those in the gas, though larger dust grains produce higher contrast features.

Regarding the gaps in Fig. \ref{fig:sigma_visc}, in general, for gas, $\gd$ at $\alpha = 10^{-4}$ and $\alpha = 10^{-3}$ is approximately 2 and 1 orders of magnitude deeper, respectively, than at $\alpha = 10^{-2}$. For dust, small grains couple well with the gas, while larger dust grains exhibit more pronounced structuring.
Specifically, at high viscosity ($\alpha = 10^{-2}$), the mm dust gap is about $80\%$ depleted. In the $\alpha = 10^{-3}$ case, a wide mm dust gap opens between 7 and 13 au, with the gap being nearly empty ($\gd < 10^{-3}$). At $\alpha = 10^{-4}$, multiple gas and dust gaps appear, which can be attributed to the secondary spiral arms excited by the planet \citep{zhu_particle_2014, bae_formation_2017, dong_multiple_2017}. The main gap forms around $\rp$, while a shallow secondary gap appears at 7 au (0.7 $\rp$), and a deep ($90\%$ depleted) secondary dust gap is located 5 au beyond $\rp$. The positions of these secondary gaps align with the findings of \citet{zhang_disk_2018}.

In terms of ring structure, three mm dust density rings are present at $\alpha = 10^{-4}$. The ring at the outer gap edge is the strongest dust trap, showing an order-of-magnitude density enhancement. Additionally, a density peak is visible at the middle of the gap, likely caused by mm dust remaining in the horseshoe orbit. This could be a transient feature, as 2000 orbits may not be sufficient for the system to reach a steady state at $\alpha = 10^{-4}$.

\begin{figure*}
\includegraphics[width=\linewidth]{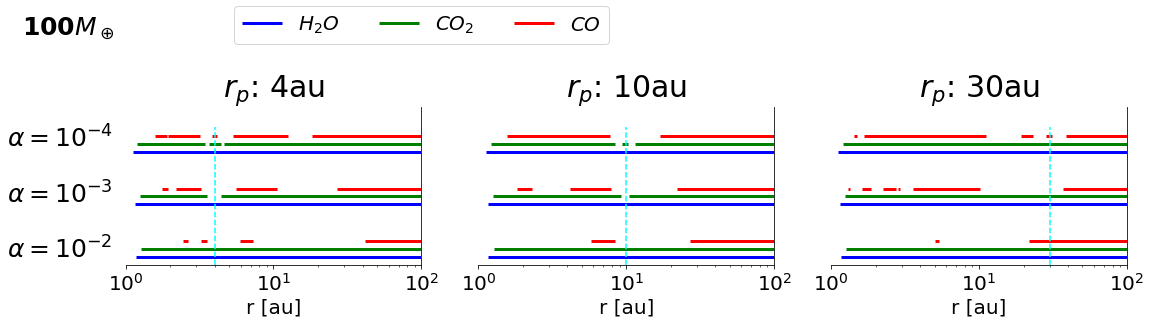} 
\centering
\caption{Iceline locations obtained from models with $\Mp=100\Me$ and viscosity of $\alpha=10^{-2}$ (bottom), $10^{-3}$ (middle) and $\alpha=10^{-4}$ (top). The vertical cyan lines mark $\rp$. 
}
\label{fig:ice_visc_100me} 
\end{figure*}

We present the results of the ice distribution for a 100 $\Me$ planet in Fig.~\ref{fig:ice_visc_100me}.
First, outside the gap region, we observe the same effects as in Fig.~\ref{fig:ice_visc_noplanet}. Lower viscosity predicts a lower $\Tmid$, causing the \ce{CO} ice region to start closer to the star.
Second, within the gap region, there is no simple correlation between viscosity values and the widths of the \ce{CO} sublimation regions due to gap opening. On one hand, lower $\alpha$ viscosity results in a wider and deeper gap, which increases $\Tmid$. On the other hand, in a disk without a planet, lower viscosity leads to a decrease in $\Tmid$. As a result, the combined effect also weakens the correlation between $\alpha$ viscosity and iceline locations and numbers.
Notably, in cases with $\alpha = 10^{-4}$, some very short bars representing \ce{CO2} or \ce{CO} ice appear within the gap region. This occurs due to mm-sized dust remaining in horseshoe orbits or forming dust clumps at the L4 and L5 points. These mm dust overdensities at $\rp$ cause a local drop in $\Tmid$, leading to the freeze-out of \ce{CO} or \ce{CO2}.

\section{Discussion}

We discuss the implications of our results on the disk temperature structure and the observability of gap temperature changes. Also, we discuss the limtis of our model.

\subsection{Rings/gaps in hydro simulations vs molecule line observations}
\label{sec:hd_vs_obs}

A deep gas gap identified in hydrodynamical (HD) simulations may not necessarily appear as a gap in molecular line observations. This discrepancy arises because HD simulations primarily model gas as \ce{H2}, which is not directly observable. Observations, such as those conducted with ALMA, trace specific molecular species. For example, CO can freeze out or be photodissociated in certain disk regions and exhibit depletion levels different from those of \ce{H2} \citep[e.g.,][]{schwarz_unlocking_2018,schwarz_unlocking_2019, krijt_co_2020}. 

Here we consider the case of \ce{CO} and ignore photodissociation for simplicity. 
In a smooth disk, where no planet-induced substructures are present, \ce{CO} remains in the gas phase at radii smaller than the \ce{CO} iceline location, $R_{co,ice}$. If a deep \ce{H2} gas gap is introduced at this location, the \ce{CO} depletion follows that of \ce{H2}, resulting in an observable gap in \ce{CO} emission. Conversely, beyond $R_{co,ice}$ in a smooth disk, \ce{CO} is expected to freeze out onto dust grains, appearing as a dark region in observations. However, if a deep gas gap forms at a radial location outside $R_{co,ice}$ (i.e., beyond where the \ce{CO} iceline would be in a smooth disk), the reduced gas density and altered thermal structure may lead to the sublimation of \ce{CO}, allowing it to return to the gas phase and forming a bright molecular ring in observations.

Thus, a deep \ce{H2} gas gap in a structured disk may manifest as a molecular ring if it is located outside the \ce{CO} iceline of a smooth disk. Conversely, an observed molecular gap at radial location inside $R_{co,ice}$ of the couterpart smooth disk may be caused either by \ce{CO} freeze-out, potentially induced by dust rings or shadowing effects, or by a true deep gas gap in the disk. This highlights the importance of carefully interpreting molecular observations in the context of disk substructure and temperature.

\subsection{Observability of planet impact on disk temperature and icelines}

Previous observations of CO and CO isotopologue icelines accompanied by \ce{N2H+} at $\sim$20–30 au in the protoplanetary disk TW Hya have been reported in \citet{qi_imaging_2013,zhang_mass_2017}. These studies show a sharp drop in CO intensity at the iceline location.

Our structured disk modeling suggests that deep gaps opened by giant planets can significantly increase the local temperature (Fig. \ref{fig:tmid_rp30}). If a gap forms outside the iceline of a smooth disk, it can sublimate volatile ices such as CO back into the gas phase. Consequently, CO abundance will be higher at the gap compared to a disk without a gap, and the CO iceline will shift outward.

The above idea is potentially testable by comparing the CO iceline locations measured by ALMA with our disk model, where we place planets of different masses into the same disk setup. If a planet-induced gap is present, CO emission should extend farther out in intensity maps. For instance, in Fig. \ref{fig:ice_dust}, in the 3$\Mj$ at 30 au case, the CO iceline moves to 50 au (top third in panel(a2)), whereas in the 10$\Me$ at 30 au case (where the planet cannot open a gap), it remains at 20 au (top third in panel(c2)). In this case, a resolution of 30 au is required to resolve the change in the CO emission area. However, if this resolution is not achievable, CO spectral line profiles can still indicate the gap temperature change through an increase in CO intensity flux.

Additionally, CO ice sublimation at the gap could leave kinematic signatures in channel maps. Simply speaking, some velocity channels should show stronger CO emission in gap regions compared to gap-free regions. However, as shown in \citet{chen_mind_2024}, both the inner and outer gap edges can exhibit strong gas velocity perturbations (on the order of $\sim$0.1 of the local Keplerian velocity). Meanwhile, spirals near the planet can also induce velocity perturbations at similar levels as those at the gap edges. These perturbations can affect the intensity of CO emission in channel maps, potentially overlapping with the effects of gap temperature changes. Therefore, accurate modeling requires 3D hydrodynamical and radiative transfer simulations to predict the kinematic signatures in the future.

\subsection{``Flickering'' icelines}
\label{sec:flickering_ice}

We have noted that complex, small-scale ice rings usually develop in the inner disk ($\lesssim 5$ to $10\,$au, e.g., see Fig.\,\ref{fig:ice_dust} and the case of $\alpha=10^{-3}$ in Fig. \ref{fig:ice_visc_noplanet}) in our simulations. This is because the disk midplane conditions end up close to the local CO sublimation temperature, which makes the presence or lack of CO ice highly sensitive to fluctuations in the hydrodynamical models.

We propose that, while this makes it hard to pin down one specific iceline location, the phenomenon may be real. This would manifest as ``flickering'' icelines, where large regions of the inner disk may irregularly fluctuate between CO freeze-out and sublimation, depending on small variations in the local disk conditions. Such a cycling of ice and gas phases in regions spanning anywhere from a small fraction of an au to $\sim1\,$au in extent may have implications for the ice and gas chemistry, as well as in the ice composition inherited by pebbles or planetesimals in these disk regions.

\subsection{Limits of our model}
\label{sec:limits}

First, our models neglect the effects of dust growth processes, including fragmentation, coagulation, and cratering. The timescales of dust growth may be comparable to those of dust drift, settling, and diffusion \citep{birnstiel_dust_2023}. Dust growth alters grain sizes, which in turn affects dust opacities, influencing heating and cooling processes and ultimately modifying the disk temperature and iceline locations. Despite this, \citet{savvidou_influence_2020} finds the temperature comparisons between the simple power-lawer grain size model \citep{mathis_size_1977} is still similar (difference less than 10\,K) to the more complex dust growth model \citet{birnstiel_dust_2011} in regions outside the first few au in an equilibrium disk. 

Second, in our iterative process, we assume that the surface-area-averaged dust temperature is equal to the gas temperature. However, this assumption is not always valid. For example, \citet{facchini_inferring_2018} show that in gap regions, gas and dust temperatures can decouple due to the reduced dust surface area. Specifically, they find that in the midplane of a deep gap, $\Tgas/\Tdust < 1$. Such lower $\Tgas$ can lead to deeper gas and dust gaps in hydrodynamical simulations.

Third, in our MCRT temperature calculations, we consider only stellar radiation. However, disk temperature can also be influenced by other factors, such as viscous heating and external radiation. Viscous heating primarily affects the disk midplane within the innermost few au. For a fixed viscosity, without gap formation, viscous heating increases $\Tmid$ in this region. However, if a gap forms at a few au, although more stellar photons can penetrate into the gap, even more viscous heat is lost. For example, \citet{broome_iceline_2023} shows $\Tmid$ within the gap (at around 3 au) form by a Jovian planet can decreases 20\% to 30\% from that of gap-free model. 
For different $\alpha$ viscosities, the temperature would generally increase with higher viscosity \citep{savvidou_influence_2020}. Therefore, it would enhance the temperature difference that we show in Fig. \ref{fig:tmid_visc_noplanet}.
Additionally, external radiation sources, such as cosmic rays and external photoevaporation, can heat the outer disk regions.

Fourth, as the shifting, multiplication, and ``flickering'' (see Section\,\ref{sec:flickering_ice}) of icelines is dependent on general properties of the spatial temperature profile and its potential closeness to the sublimation temperature of specific volatile species, we do not expect those findings to fundamentally change with further improvements to the simulations (e.g., using a higher number of photon packages in MCRT). However, properties such as the absolute location or number of icelines for a given volatile chemical species may change somewhat.

\section{Conclusions}

We iterate the $\hd$ and $\rt$ simulations to study the planet-induced gas and dust substructures in disks and how these structures can affect the disk temperature structure. We compare our new models, including gas and multiple dust ($\md$), with our old models with gas only ($\mg$). In addition, we investigate how different turbulent viscosities can influence our $\md$. Here are our main findings:

(1) Regarding density structure, compared to $\mg$, $\md$ predicts shallower gas gaps due to higher temperatures in the gap region. For instance, a 100$\Me$ planet at 10 au in $\md$  opens a gas ($\gd \simeq 5\times10^{-2}$ and dust ($\gd < 10^{-6}$) gap, while the gas gap in $\mg$ is about 1.5 times deeper (Fig. \ref{fig:3plot_100me10au}).  

(2) For the temperature at substructures, gaps or dust rings, like $\mg$, $\md$ also finds that a giant planet \bt{(e.g. $3\Mj$ at 30 au)} can increase the midplane temperature by a few tens of K \bt{(from 30 to 60\,K)} in the gap region (Fig. \ref{fig:tmid_rp30}).
In $\md$, a 3$\Mj$ planet at 30 au forms a mm dust ring next to the outer gap edge, cooling $\Tmid$ by several K and creating a potential freeze-out zone (Figs.~\ref{fig:dust_trap}).

(3) Across the whole disk radius, the midplane temperature difference is not significant between $\mg$ and $\md$. Combining the sublimation temperature of volatiles, we find that the ice distribution of \ce{H2O}, \ce{CO2}, and \ce{CO} is similar between these two models (Fig. \ref{fig:ice_dust}).

(4) \bt{In non-planet disks in $\md$, decreasing $\alpha$ viscosity (from $10^{-2}$ to $10^{-4}$) weakens turbulent mixing and enhances dust settling, cooling the midplane by $\sim$10\,K ($\sim 25\%$)}. This shifts the \ce{CO} iceline (outside the self-shadowing region) inward, from 40 au ($\alpha=10^{-2}$) to 20 au (Fig.~\ref{fig:tmid_visc_noplanet}).

(5) With a planet in $\md$, low viscosity allows deeper gaps and stronger heating, complicating the relation between viscosity and gap temperature. This complexity is reflected in the width of CO sublimation regions (Fig.~\ref{fig:ice_visc_100me}).

(6) Planet-induced gaps can heat the disk locally, sublimating CO ice and pushing the CO iceline outward. This may be detectable with ALMA via CO intensity maps or spectral lines. However, for channel maps, velocity perturbations at gap edges and spirals may mimic thermal effects, requiring detailed 3D modeling to disentangle them.

\section*{Acknowledgements}
We are thankful to the referee for the constructive report. KC acknowledges support by UCL Dean's Prize and China Scholarship Council.  PP acknowledges funding from the UK Research and Innovation (UKRI) under the UK government’s Horizon Europe funding guarantee from ERC (under grant agreement No 101076489). MK gratefully acknowledges funding from the European Union's Horizon Europe research and innovation programme under grant agreement No. 101079231 (EXOHOST), and from UK Research and Innovation (UKRI) under the UK government’s Horizon Europe funding guarantee (grant number 10051045).


\section*{Data Availability}
Data from our numerical models are available on reasonable request to the corresponding author.
The $\fargo$ code is publicly available from
\href{https://fargo3d.bitbucket.io/download.html}{https://fargo3d.bitbucket.io/download.html}.
The $\radmc$ code is available from \href{https://www.ita.uni-heidelberg.de/~dullemond/software/radmc-3d/}{https://www.ita.uni-heidelberg.de/~dullemond/software/radmc-3d/}.



\bibliographystyle{mnras}
\bibliography{zotero.bib}

\begin{thebibliography}{}
\makeatletter
\relax
\def\mn@urlcharsother{\let\do\@makeother \do\$\do\&\do\#\do\^\do\_\do\%\do\~}
\def\mn@doi{\begingroup\mn@urlcharsother \@ifnextchar [ {\mn@doi@} {\mn@doi@[]}}
\def\mn@doi@[#1]#2{\def\@tempa{#1}\ifx\@tempa\@empty \href {http://dx.doi.org/#2} {doi:#2}\else \href {http://dx.doi.org/#2} {#1}\fi \endgroup}
\def\mn@eprint#1#2{\mn@eprint@#1:#2::\@nil}
\def\mn@eprint@arXiv#1{\href {http://arxiv.org/abs/#1} {{\tt arXiv:#1}}}
\def\mn@eprint@dblp#1{\href {http://dblp.uni-trier.de/rec/bibtex/#1.xml} {dblp:#1}}
\def\mn@eprint@#1:#2:#3:#4\@nil{\def\@tempa {#1}\def\@tempb {#2}\def\@tempc {#3}\ifx \@tempc \@empty \let \@tempc \@tempb \let \@tempb \@tempa \fi \ifx \@tempb \@empty \def\@tempb {arXiv}\fi \@ifundefined {mn@eprint@\@tempb}{\@tempb:\@tempc}{\expandafter \expandafter \csname mn@eprint@\@tempb\endcsname \expandafter{\@tempc}}}

\bibitem[\protect\citeauthoryear{Alarcón, Teague, Zhang, Bergin  \& Barraza-Alfaro}{Alarcón et~al.}{2020}]{alarcon_chemical_2020}
Alarcón F.,  Teague R.,  Zhang K.,  Bergin E.~A.,   Barraza-Alfaro M.,  2020, \mn@doi [The Astrophysical Journal] {10.3847/1538-4357/abc1d6}, 905, 68

\bibitem[\protect\citeauthoryear{Andrews et~al.,}{Andrews et~al.}{2018}]{andrews_disk_2018}
Andrews S.~M.,  et~al., 2018, \mn@doi [The Astrophysical Journal] {10.3847/2041-8213/aaf741}, 869, L41

\bibitem[\protect\citeauthoryear{Bae}{Bae}{2017}]{bae_formation_2017}
Bae J.,  2017, \mn@doi [The Astrophysical Journal] {https://doi.org/10.3847/1538-4357/aa9705}, p.~10

\bibitem[\protect\citeauthoryear{Birnstiel}{Birnstiel}{2023}]{birnstiel_dust_2023}
Birnstiel T.,  2023, Dust growth and evolution in protoplanetary disks, \url {http://arxiv.org/abs/2312.13287}

\bibitem[\protect\citeauthoryear{Birnstiel, Ormel  \& Dullemond}{Birnstiel et~al.}{2011}]{birnstiel_dust_2011}
Birnstiel T.,  Ormel C.~W.,   Dullemond C.~P.,  2011, \mn@doi [Astronomy \& Astrophysics] {10.1051/0004-6361/201015228}, 525, A11

\bibitem[\protect\citeauthoryear{Broome, Kama, Booth  \& Shorttle}{Broome et~al.}{2023}]{broome_iceline_2023}
Broome M.,  Kama M.,  Booth R.,   Shorttle O.,  2023, \mn@doi [Monthly Notices of the Royal Astronomical Society] {10.1093/mnras/stad1159}, 522, 3378

\bibitem[\protect\citeauthoryear{Calahan et~al.,}{Calahan et~al.}{2021}]{calahan_tw_2021}
Calahan J.~K.,  et~al., 2021, \mn@doi [The Astrophysical Journal] {10.3847/1538-4357/abd255}, 908, 8

\bibitem[\protect\citeauthoryear{Chen \& Dong}{Chen \& Dong}{2024}]{chen_mind_2024}
Chen K.,  Dong R.,  2024, \mn@doi [The Astrophysical Journal] {10.3847/1538-4357/ad83d0}, 976, 49

\bibitem[\protect\citeauthoryear{Chen, Kama, Pinilla  \& Keyte}{Chen et~al.}{2023}]{chen_planet_2023}
Chen K.,  Kama M.,  Pinilla P.,   Keyte L.,  2023, \mn@doi [Monthly Notices of the Royal Astronomical Society] {10.1093/mnras/stad3247}, 527, 2049

\bibitem[\protect\citeauthoryear{Crida, Morbidelli  \& Masset}{Crida et~al.}{2006}]{crida_width_2006}
Crida A.,  Morbidelli A.,   Masset F.,  2006, \mn@doi [Icarus] {10.1016/j.icarus.2005.10.007}, 181, 587

\bibitem[\protect\citeauthoryear{Dominik, Min  \& Tazaki}{Dominik et~al.}{2021}]{dominik_optool_2021}
Dominik C.,  Min M.,   Tazaki R.,  2021, Astrophysics Source Code Library, p. ascl:2104.010

\bibitem[\protect\citeauthoryear{Dong, Li, Chiang  \& Li}{Dong et~al.}{2017}]{dong_multiple_2017}
Dong R.,  Li S.,  Chiang E.,   Li H.,  2017, \mn@doi [The Astrophysical Journal] {10.3847/1538-4357/aa72f2}, 843, 127

\bibitem[\protect\citeauthoryear{Duffell}{Duffell}{2020}]{duffell_empirically_2020}
Duffell P.~C.,  2020, \mn@doi [The Astrophysical Journal] {10.3847/1538-4357/ab5b0f}, 889, 16

\bibitem[\protect\citeauthoryear{Dullemond \& Dominik}{Dullemond \& Dominik}{2004}]{dullemond_effect_2004}
Dullemond C.~P.,  Dominik C.,  2004, \mn@doi [Astronomy \& Astrophysics] {10.1051/0004-6361:20040284}, 421, 1075

\bibitem[\protect\citeauthoryear{Dullemond, Juhasz, Pohl, Sereshti, Shetty, Peters, Commercon  \& Flock}{Dullemond et~al.}{2012}]{dullemond_radmc-3d_2012}
Dullemond C.~P.,  Juhasz A.,  Pohl A.,  Sereshti F.,  Shetty R.,  Peters T.,  Commercon B.,   Flock M.,  2012, Astrophysics Source Code Library, p. ascl:1202.015

\bibitem[\protect\citeauthoryear{Dullemond et~al.,}{Dullemond et~al.}{2018}]{dullemond_disk_2018}
Dullemond C.~P.,  et~al., 2018, \mn@doi [The Astrophysical Journal] {10.3847/2041-8213/aaf742}, 869, L46

\bibitem[\protect\citeauthoryear{Facchini, Birnstiel, Bruderer  \& van Dishoeck}{Facchini et~al.}{2017}]{facchini_different_2017}
Facchini S.,  Birnstiel T.,  Bruderer S.,   van Dishoeck E.~F.,  2017, \mn@doi [Astronomy \& Astrophysics] {10.1051/0004-6361/201630329}, 605, A16

\bibitem[\protect\citeauthoryear{Facchini, Pinilla, van Dishoeck  \& de Juan~Ovelar}{Facchini et~al.}{2018}]{facchini_inferring_2018}
Facchini S.,  Pinilla P.,  van Dishoeck E.~F.,   de Juan~Ovelar M.,  2018, \mn@doi [Astronomy \& Astrophysics] {10.1051/0004-6361/201731390}, 612, A104

\bibitem[\protect\citeauthoryear{Fedele, Van~Dishoeck, Kama, Bruderer  \& Hogerheijde}{Fedele et~al.}{2016}]{fedele_probing_2016}
Fedele D.,  Van~Dishoeck E.~F.,  Kama M.,  Bruderer S.,   Hogerheijde M.~R.,  2016, \mn@doi [Astronomy \& Astrophysics] {10.1051/0004-6361/201526948}, 591, A95

\bibitem[\protect\citeauthoryear{Flaherty, Hughes, Rosenfeld, Andrews, Chiang, Simon, Kerzner  \& Wilner}{Flaherty et~al.}{2015}]{flaherty_weak_2015}
Flaherty K.~M.,  Hughes A.~M.,  Rosenfeld K.~A.,  Andrews S.~M.,  Chiang E.,  Simon J.~B.,  Kerzner S.,   Wilner D.~J.,  2015, \mn@doi [The Astrophysical Journal] {10.1088/0004-637X/813/2/99}, 813, 99

\bibitem[\protect\citeauthoryear{Flaherty, Hughes, Teague, Simon, Andrews  \& Wilner}{Flaherty et~al.}{2018}]{flaherty_turbulence_2018}
Flaherty K.,  Hughes A.,  Teague R.,  Simon J.,  Andrews S.,   Wilner D.,  2018, \mn@doi [The Astrophysical Journal] {10.3847/1538-4357/aab615}, 856, 117

\bibitem[\protect\citeauthoryear{Flaherty et~al.,}{Flaherty et~al.}{2020}]{flaherty_measuring_2020}
Flaherty K.,  et~al., 2020, \mn@doi [The Astrophysical Journal] {10.3847/1538-4357/ab8cc5}, 895, 109

\bibitem[\protect\citeauthoryear{Fromang \& Nelson}{Fromang \& Nelson}{2009}]{fromang_global_2009}
Fromang S.,  Nelson R.~P.,  2009, \mn@doi [Astronomy \& Astrophysics] {10.1051/0004-6361/200811220}, 496, 597

\bibitem[\protect\citeauthoryear{Fung, Shi  \& Chiang}{Fung et~al.}{2014}]{fung_how_2014}
Fung J.,  Shi J.-M.,   Chiang E.,  2014, \mn@doi [The Astrophysical Journal] {10.1088/0004-637X/782/2/88}, 782, 88

\bibitem[\protect\citeauthoryear{Hollenbach, Kaufman, Bergin  \& Melnick}{Hollenbach et~al.}{2009}]{hollenbach_water_2009}
Hollenbach D.,  Kaufman M.~J.,  Bergin E.~A.,   Melnick G.~J.,  2009, \mn@doi [The Astrophysical Journal] {10.1088/0004-637X/690/2/1497}, 690, 1497

\bibitem[\protect\citeauthoryear{Huang et~al.,}{Huang et~al.}{2018}]{huang_disk_2018}
Huang J.,  et~al., 2018, \mn@doi [The Astrophysical Journal] {10.3847/2041-8213/aaf740}, 869, L42

\bibitem[\protect\citeauthoryear{Kanagawa, Muto, Tanaka, Tanigawa, Takeuchi, Tsukagoshi  \& Momose}{Kanagawa et~al.}{2015}]{kanagawa_mass_2015}
Kanagawa K.~D.,  Muto T.,  Tanaka H.,  Tanigawa T.,  Takeuchi T.,  Tsukagoshi T.,   Momose M.,  2015, \mn@doi [The Astrophysical Journal] {10.1088/2041-8205/806/1/L15}, 806, L15

\bibitem[\protect\citeauthoryear{Krijt, Bosman, Zhang, Schwarz, Ciesla  \& Bergin}{Krijt et~al.}{2020}]{krijt_co_2020}
Krijt S.,  Bosman A.~D.,  Zhang K.,  Schwarz K.~R.,  Ciesla F.~J.,   Bergin E.~A.,  2020, \mn@doi [The Astrophysical Journal] {10.3847/1538-4357/aba75d}, 899, 134

\bibitem[\protect\citeauthoryear{Law et~al.,}{Law et~al.}{2021a}]{lawMoleculesALMAPlanetforming2021}
Law C.~J.,  et~al., 2021a, \mn@doi [The Astrophysical Journal Supplement Series] {10.3847/1538-4365/ac1434}, 257, 3

\bibitem[\protect\citeauthoryear{Law et~al.,}{Law et~al.}{2021b}]{lawMoleculesALMAPlanetforming2021a}
Law C.~J.,  et~al., 2021b, \mn@doi [The Astrophysical Journal Supplement Series] {10.3847/1538-4365/ac1439}, 257, 4

\bibitem[\protect\citeauthoryear{Law et~al.,}{Law et~al.}{2024}]{law_mapping_2024}
Law C.~J.,  et~al., 2024, Mapping the {Vertical} {Gas} {Structure} of the {Planet}-hosting {PDS} 70 {Disk}, \url {http://arxiv.org/abs/2401.03018}

\bibitem[\protect\citeauthoryear{Leemker et~al.,}{Leemker et~al.}{2022}]{leemker_gas_2022}
Leemker M.,  et~al., 2022, arXiv:2204.03666 [astro-ph]

\bibitem[\protect\citeauthoryear{Long et~al.,}{Long et~al.}{2018}]{long_gaps_2018}
Long F.,  et~al., 2018, \mn@doi [The Astrophysical Journal] {10.3847/1538-4357/aae8e1}, 869, 17

\bibitem[\protect\citeauthoryear{Mathis, Rumpl  \& Nordsieck}{Mathis et~al.}{1977}]{mathis_size_1977}
Mathis J.~S.,  Rumpl W.,   Nordsieck K.~H.,  1977, \mn@doi [The Astrophysical Journal] {10.1086/155591}, 217, 425

\bibitem[\protect\citeauthoryear{Pinilla, Birnstiel, Ricci, Dullemond, Uribe, Testi  \& Natta}{Pinilla et~al.}{2012a}]{pinilla_trapping_2012}
Pinilla P.,  Birnstiel T.,  Ricci L.,  Dullemond C.~P.,  Uribe A.~L.,  Testi L.,   Natta A.,  2012a, \mn@doi [Astronomy \& Astrophysics] {10.1051/0004-6361/201118204}, 538, A114

\bibitem[\protect\citeauthoryear{Pinilla, Benisty  \& Birnstiel}{Pinilla et~al.}{2012b}]{pinilla_ring_2012}
Pinilla P.,  Benisty M.,   Birnstiel T.,  2012b, \mn@doi [Astronomy \& Astrophysics] {10.1051/0004-6361/201219315}, 545, A81

\bibitem[\protect\citeauthoryear{Pinilla, Pohl, Stammler  \& Birnstiel}{Pinilla et~al.}{2017}]{pinilla_dust_2017}
Pinilla P.,  Pohl A.,  Stammler S.~M.,   Birnstiel T.,  2017, \mn@doi [The Astrophysical Journal] {10.3847/1538-4357/aa7edb}, 845, 68

\bibitem[\protect\citeauthoryear{Pinte, Dent, Ménard, Hales, Hill, Cortes  \& Gregorio-Monsalvo}{Pinte et~al.}{2016}]{pinte_dust_2016}
Pinte C.,  Dent W. R.~F.,  Ménard F.,  Hales A.,  Hill T.,  Cortes P.,   Gregorio-Monsalvo I.~D.,  2016, \mn@doi [The Astrophysical Journal] {10.3847/0004-637X/816/1/25}, 816, 25

\bibitem[\protect\citeauthoryear{Pyerin, Delage, Kurtovic, Gárate, Henning  \& Pinilla}{Pyerin et~al.}{2021}]{pyerin_constraining_2021}
Pyerin M.~A.,  Delage T.~N.,  Kurtovic N.~T.,  Gárate M.,  Henning T.,   Pinilla P.,  2021, arXiv:2110.03373 [astro-ph]

\bibitem[\protect\citeauthoryear{Qi et~al.,}{Qi et~al.}{2013}]{qi_imaging_2013}
Qi C.,  et~al., 2013, \mn@doi [Science] {10.1126/science.1239560}, 341, 630

\bibitem[\protect\citeauthoryear{Rosotti, Juhasz, Booth  \& Clarke}{Rosotti et~al.}{2016}]{rosotti_minimum_2016}
Rosotti G.~P.,  Juhasz A.,  Booth R.~A.,   Clarke C.~J.,  2016, \mn@doi [Monthly Notices of the Royal Astronomical Society] {10.1093/mnras/stw691}, 459, 2790

\bibitem[\protect\citeauthoryear{Savvidou, Bitsch  \& Lambrechts}{Savvidou et~al.}{2020}]{savvidou_influence_2020}
Savvidou S.,  Bitsch B.,   Lambrechts M.,  2020, \mn@doi [Astronomy \& Astrophysics] {10.1051/0004-6361/201936576}, 640, A63

\bibitem[\protect\citeauthoryear{Schoonenberg \& Ormel}{Schoonenberg \& Ormel}{2017}]{schoonenberg_planetesimal_2017}
Schoonenberg D.,  Ormel C.~W.,  2017, \mn@doi [Astronomy \& Astrophysics] {10.1051/0004-6361/201630013}, 602, A21

\bibitem[\protect\citeauthoryear{Schwarz, Bergin, Cleeves, Zhang, Öberg, Blake  \& Anderson}{Schwarz et~al.}{2018}]{schwarz_unlocking_2018}
Schwarz K.~R.,  Bergin E.~A.,  Cleeves L.~I.,  Zhang K.,  Öberg K.~I.,  Blake G.~A.,   Anderson D.,  2018, \mn@doi [The Astrophysical Journal] {10.3847/1538-4357/aaae08}, 856, 85

\bibitem[\protect\citeauthoryear{Schwarz, Bergin, Cleeves, Zhang, Öberg, Blake  \& Anderson}{Schwarz et~al.}{2019}]{schwarz_unlocking_2019}
Schwarz K.~R.,  Bergin E.~A.,  Cleeves L.~I.,  Zhang K.,  Öberg K.~I.,  Blake G.~A.,   Anderson D.~E.,  2019, \mn@doi [The Astrophysical Journal] {10.3847/1538-4357/ab1c5e}, 877, 131

\bibitem[\protect\citeauthoryear{Teague et~al.,}{Teague et~al.}{2018}]{teague_temperature_2018}
Teague R.,  et~al., 2018, \mn@doi [The Astrophysical Journal] {10.3847/1538-4357/aad80e}, 864, 133

\bibitem[\protect\citeauthoryear{Turner, Choukroun, Castillo-Rogez  \& Bryden}{Turner et~al.}{2012}]{turner_hot_2012}
Turner N.~J.,  Choukroun M.,  Castillo-Rogez J.,   Bryden G.,  2012, \mn@doi [The Astrophysical Journal] {10.1088/0004-637X/748/2/92}, 748, 92

\bibitem[\protect\citeauthoryear{Weber, Pérez, Benítez-Llambay, Gressel, Casassus  \& Krapp}{Weber et~al.}{2019}]{weber_predicting_2019}
Weber P.,  Pérez S.,  Benítez-Llambay P.,  Gressel O.,  Casassus S.,   Krapp L.,  2019, \mn@doi [The Astrophysical Journal] {10.3847/1538-4357/ab412f}, 884, 178

\bibitem[\protect\citeauthoryear{Weidenschilling}{Weidenschilling}{1977}]{weidenschilling_aerodynamics_1977}
Weidenschilling S.~J.,  1977, \mn@doi [Monthly Notices of the Royal Astronomical Society] {10.1093/mnras/180.2.57}, 180, 57

\bibitem[\protect\citeauthoryear{Zhang, Blake  \& Bergin}{Zhang et~al.}{2015}]{zhang_evidence_2015}
Zhang K.,  Blake G.~A.,   Bergin E.~A.,  2015, \mn@doi [The Astrophysical Journal] {10.1088/2041-8205/806/1/L7}, 806, L7

\bibitem[\protect\citeauthoryear{Zhang, Bergin, Blake, Cleeves  \& Schwarz}{Zhang et~al.}{2017}]{zhang_mass_2017}
Zhang K.,  Bergin E.~A.,  Blake G.~A.,  Cleeves L.~I.,   Schwarz K.~R.,  2017, \mn@doi [Nature Astronomy] {10.1038/s41550-017-0130}, 1, 0130

\bibitem[\protect\citeauthoryear{Zhang et~al.,}{Zhang et~al.}{2018}]{zhang_disk_2018}
Zhang S.,  et~al., 2018, \mn@doi [The Astrophysical Journal] {10.3847/2041-8213/aaf744}, 869, L47

\bibitem[\protect\citeauthoryear{Zhang, Hu, Zhu  \& Bae}{Zhang et~al.}{2021}]{zhang_self-consistent_2021}
Zhang S.,  Hu X.,  Zhu Z.,   Bae J.,  2021, arXiv:2110.00858 [astro-ph]

\bibitem[\protect\citeauthoryear{Zhu, Stone, Rafikov  \& Bai}{Zhu et~al.}{2014}]{zhu_particle_2014}
Zhu Z.,  Stone J.~M.,  Rafikov R.~R.,   Bai X.-n.,  2014, \mn@doi [The Astrophysical Journal] {10.1088/0004-637X/785/2/122}, 785, 122

\bibitem[\protect\citeauthoryear{Öberg, Murray-Clay  \& Bergin}{Öberg et~al.}{2011}]{oberg_effects_2011}
Öberg K.~I.,  Murray-Clay R.,   Bergin E.~A.,  2011, \mn@doi [The Astrophysical Journal] {10.1088/2041-8205/743/1/L16}, 743, L16

\bibitem[\protect\citeauthoryear{Öberg et~al.,}{Öberg et~al.}{2021}]{oberg_molecules_2021}
Öberg K.~I.,  et~al., 2021, \mn@doi [The Astrophysical Journal Supplement Series] {10.3847/1538-4365/ac1432}, 257, 1

\makeatother
\end{thebibliography}



\appendix

\section{Temperature comparison between Model G and Model D }
\label{sec:app_Tdiff_modelg_modeld}

\begin{figure*}
\includegraphics[width=\linewidth]{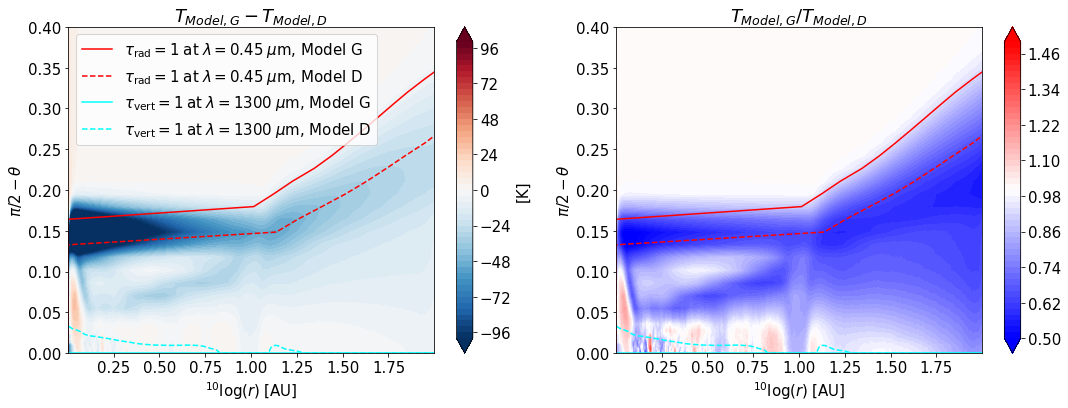}
\centering
\caption{Temperature difference (left) and ratio (right) between $\mg$ and $\md$ in Fig.~\ref{fig:tau1_wi_wt_dust}. Their corresponding $\taur=1$ surfaces at 0.45 $\mu$m and $\tauv=1$ surfaces at 1.3 mm are shown in each panel. 
}
\label{fig:Tdiff_modelg_modeld}
\end{figure*}

Figure \ref{fig:Tdiff_modelg_modeld} shows the temperature comparison between $\mg$ and $\md$ in Fig.~\ref{fig:tau1_wi_wt_dust}. In most regions except the midplane, the temperature of $\mg$ is about 30\% (or even more) cooler than that of $\md$. Especially near the disk surface, the temperature deviation can be more than 100 K or about 50\% difference. This is because the $\taur=1$ surfaces at 0.45 $\mu$m in $\mg$ is about 30\% higher than that in $\md$, which makes the heating weaker and temperauture lower. At regions near the disk midplane, especially $r<10$ au, the temperature of $\mg$ get relatively close to that of $\md$, with a difference of about 15\%.

\section{Temperature comparison between $\alpha=10^{-2}$ and $\alpha=10^{-4}$}
\label{sec:app_Tdiff_visc}

\begin{figure*}
\includegraphics[width=\linewidth]{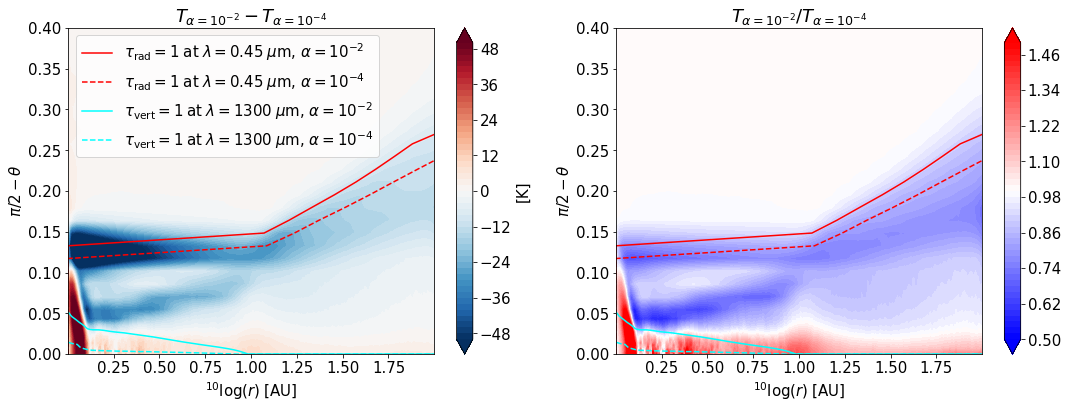}
\centering
\caption{Temperature difference (left) and ratio (right) between $\alpha=10^{-2}$ and $\alpha=10^{-4}$ in Fig.~\ref{fig:tau1_visc_noplanet}. Their corresponding $\taur=1$ surfaces at 0.45 $\mu$m and $\tauv=1$ surfaces at 1.3 mm are shown in each panel. 
}
\label{fig:Tdiff_visc}
\end{figure*}
Figure \ref{fig:Tdiff_visc} shows the temperature comparisons between $\alpha=10^{-2}$ and $\alpha=10^{-4}$ in Fig.~\ref{fig:tau1_visc_noplanet}.



\bsp	
\label{lastpage}
\end{document}